\documentclass[amsmath,amssymb,aip,jcp,citeautoscript,superscriptaddress,twocolumn,
10pt]{revtex4-2}

\usepackage{bm}
\usepackage{graphicx}
\usepackage{dcolumn}
\usepackage[T1]{fontenc} 
\usepackage[colorlinks,linkcolor=red,anchorcolor=green,
citecolor=blue,breaklinks]{hyperref}

\usepackage{float}
\usepackage{mathrsfs}
\usepackage{color}
\usepackage[normalem]{ulem}
\usepackage{braket}

\newcommand{\bR}{\mathbf{R}}
\newcommand{\bd}{\mathbf{d}}

\newcommand{\bM}{\mathbf{M}}
\newcommand{\bP}{\mathbf{P}}

\newcommand{\bu}{\boldsymbol{\mu}}

\newcommand{\blam}{\boldsymbol{\lambda}}

\newcommand{\hbu}{\hat{\boldsymbol{\mu}}}

\newcommand{\ha}{\hat{a}}

\newcommand{\hc}{\hat{c}}

\newcommand{\hH}{\hat{H}}

\newcommand{\hT}{\hat{T}}
\newcommand{\hV}{\hat{V}}
\newcommand{\hp}{\hat{p}}
\newcommand{\hq}{\hat{q}}

\usepackage{multirow}

\graphicspath{{figures/}}
\begin{document}


\title{Semiclassical Nonadiabatic Molecular Dynamics for Molecular Exciton-Polaritons}

\author{Xinyang Li}
\affiliation{Theoretical Division, Los Alamos National Laboratory, Los Alamos, New Mexico, 87545, USA}

\author{Sergei Tretiak}
\affiliation{Theoretical Division, Los Alamos National Laboratory, Los Alamos, New Mexico, 87545, USA}
\affiliation{Center for Integrated Nanotechnologies, Los Alamos National Laboratory, Los Alamos, New Mexico 87545, USA}

\author{Yu Zhang}
\email{zhy@lanl.gov}
\affiliation{Theoretical Division, Los Alamos National Laboratory, Los Alamos, New Mexico, 87545, USA}

\date{\today}

\begin{abstract}
When the interaction between a molecular system and confined light modes in an optical or plasmonic cavity is strong enough to overcome the dissipative process, hybrid light-matter states (polaritons) emerge as the fundamental excitations in the system. Mixing the light and matter characters modifies molecules' photophysical and photochemical properties. It was reported that polaritonic states can be employed to control photochemical reactions, charge and energy transfer, and other processes. In addition, according to recent studies, vibrational strong coupling can be employed to enhance thermally activated chemical reactions resonantly. This work adopts a coherent state-based many-body state as the basis function to expand the light-matter Hamiltonian. The corresponding nonadiabatic Molecular Dynamics scheme is derived and implemented in the NEXMD package based on the semiclassical Ab Initio Multiple Cloning (AIMC) protocol. The scheme is demonstrated via a model system, pyridine, to demonstrate its validity. Our numerical results show that coherence and decoherence processes are distinguishable during the initial relaxation time in the AIMC simulations, while such dynamics are missing in the mixed quantum-classical surface hopping approach. Additionally, our results show that the light-matter coupling has smaller impact on the dynamics in surface hopping simulations. In the AIMC simulations, on the other hand, the relaxation time is increased, leading to a slower internal conversion and decreased equilibrium population build-up in the lowest electronically excited state. This highlights the importance of proper treatment of coherence in simulating exciton polaritons dynamics.
\end{abstract}

\maketitle

\section{Introduction}\label{seq:intro}
When molecular excitations hybridize with an optical field, new quasiparticles called molecular polaritons are formed due to the complex interactions among photons, electrons, and nuclei. In recent years, breakthroughs have been made to achieve strong couplings in molecular systems, and it has been shown that the strong light-matter interaction in molecular polaritons can modify the potential energy surfaces (PESs) of molecules.~\cite{Hutchison2012ACIE, GarciaVidal2021} This immediately opens up new possibilities for controlling and manipulating chemical reactions. As such, polariton chemistry, aiming to study and utilize this type of control, has become a hot topic of both experimental~\cite{Wang2014N, Zeng2016NL, Baieva2017AP, Pandya2021NC} and theoretical realms~\cite{acs7b00610,jctc6b01126,photonics7b00916, Galego2016,acsphotonics7b00305,jpclett8b02032, Weight2023pccp} in the past few years. In addition, recent developments have also achieved strong couplings in the molecular ground state, so-called vibrational strong coupling (VSC).~\cite{Thomas2019S} Experimental and theoretical works have shown that thermally-activated chemical reactions can be resonantly suppressed~\cite {Thomas2019S} or enhanced~\cite{Joelvsc2019} via VSC. 

Modeling of molecular polaritons is a theoretically complicated and computationally challenging task. The dynamical interplay among correlated electronic, photonic, and nuclear degrees of freedom (DOFs) across different time and length scales needs to be addressed accurately and on an equal footing basis.~\cite{Galego2015PRX, Flick2017PNAS} Though there are extensive recent developments in understanding the nature of polaritonic states, ranging from models systems~\cite{GonzalezTudela2013PRL, Mazza2013PRB, Spano2015JCP} and ab initio QED methods,~\cite{Galego2015PRX, Herrera2016PRL,photonics7b00680, Flick2017PNAS, Flick15285, Li:arxiv2023vt, Weight:2023qmc, Weight:2023qmc, foley:2023CPR} to machine learning (ML) algorithms~\cite{Hu2023, Li2024}, rigorous treatment of a molecular QED system would benefit from further developments.

Apart from several works where the strong coupling is achieved on a few or one-molecule levels, most experimental works involve collective coupling, where many molecules ($> 10^6$) are often coupled to the cavity modes simultaneously.~\cite{Thomas2019S, Joelvsc2019, Pino2015} As a result, including a substantial number of molecules in the simulation is crucial. When many molecules are coupled to the cavity modes, the polariton states naturally delocalize over these molecules. At the same time, there is disorder-induced localization~\cite{Timmer:2023wn}. These two effects compete with each other, underscoring the importance of a rigorous decoherence treatment in light-matter interactions. Therefore, employing a theory that correctly addresses nonadiabatic effects inside a cavity is crucial to accurately describe photochemistry under strong light-matter interactions. Indeed, there have been previous studies~\cite{jpclett6b00864,jcp4941053, C6FD00095A, Kowalewski3278,jpca7b11833,jpclett9b01599} on incorporating strong light-matter interactions in nonadiabatic molecular dynamics (NAMD). in the presence, the overwhelming computational cost makes it impractical to model any molecular system beyond a few PESs and DOFs.

Mixed quantum-classical (MQC) algorithms, such as trajectory surface hopping (TSH)~\cite{Tully1971, Tully1990} and Ehrenfest dynamics,~\cite{Ehrenfest1927} can be numerically efficient approaches in modeling polariton chemistry. However, neither method has a proper treatment of decoherence.~\cite{Nelson2020CR} Alternatively, multiconfigurational semiclassical methods, such as ab initio multiple cloning (AIMC)~\cite{Makhov2014JCP, Freixas2018PCCP, Song2021JCTC}, and ab initio multiple spawning (AIMS)~\cite{BenNun1998, Kim_2015, Rana2023JPCA}, have been developed to mitigate the problems. Here, multiple coupled Gaussian basis functions are propagated simultaneously, and the bifurcation of the nuclear wavepackets is allowed. AIMC naturally introduces decoherence and nuclear quantum effects, such as quantum-mechanical delocalization. Previous studies have shown that this cloning technique can guarantee an exact solution to the time-dependent Schr\"odinger equation at the complete basis set limit.~\cite{Symonds2018JCP, Curchod2018} By building on these strengths, this work extends the AIMC method to study polariton dynamics problems. This emerging approach allows us to explore the intricate interactions and quantum effects that arise in these systems with enhanced accuracy and insight.

The paper is organized as follows. First, we briefly overview the QED Hamiltonian in Sec.~\ref{sec:pfs} and the AIMC methods in Sec.~\ref{sec:aimc}. Then, numerical implementations and applications are presented in Sec.~\ref{sec:numerical}. Here, we model a single pyridine molecule in the gas phase. Finally, our findings are summarized in Sec.~\ref{summary}.

\section{AIMC method for polariton chemistry}
\subsection{Molecular Hamiltonian in the presence of cavity}
\label{sec:pfs}
The system of interest is a molecular system embedded in a cavity. The size of the molecule is considered to be small compared to the wavelength of the cavity model. Consequently, the electric field can be assumed to be uniform across the molecular length. Under the long wavelength approximation, the molecule-cavity hybrid system can be represented by a non-relativistic Pauli-Fierz Hamiltonian in the dipole approximation,~\cite{Flick2017PNAS, Vendrell2018CP, Schaefer2018PRA}
\begin{align}
  \label{eqn:eqham}
  \hH_{PF}=&\hH_m+\hH_c+\hH_{mc}.
\end{align}
where $\hH_m$ is the Hamiltonian of the molecule
\begin{align}\label{eqhm}
  \hH_m =& \hT_m + \hV_m \nonumber \\
    =& \sum_{IJ}\frac{1}{2\bM}\left(\hat{\bP}\delta_{IJ}-i\hbar d_{IJ}\right)^2 \hc^\dag_I \hc_J + \sum_I V_I(\bR)\hc^\dag_I\hc_I,
\end{align}
where $I$ and $J$ are indices of adiabatic states. $c_I^\dagger$ ($c_I$) is the creation (annihilation) operator for electronic excitations for state $I$. Planks constant $\hbar=1$ is assumed throughout the manuscript. $\hT_m$ is the kinetic energy operator of the nuclei with $\bM$ being the mass of nuclei. $d_{IJ}$ are the nonadiabatic couplings (NACs) between the adiabatic states. $\hV_m$ is the potential energy operator, and $V_I(\bR)$ represents the potential energy surface (PES) for the adiabatic state $I$. Within the mixed quantum-classical (e.g., TSH) and semiclassical (e.g., AIMC) molecular dynamics, the kinetics of nuclei are described by the classical Newton's equation~\cite{Nelson2020CR}, and the electronic wavefunctions are propagated quantum-mechanically, where the NAC induces the transitions between states. This work calculates the molecular properties using the semiempirical quantum mechanical (SQM) method implemented in the NEXMD package. See further details on the code in the previous articles~\cite{Malone2020JCTC, Freixas2023JCTC}.

The Hamiltonian of the cavity photon mode is
\begin{align}
  \hH_c=\sum_\alpha\omega_\alpha\left(\ha^\dag_\alpha\ha_\alpha+\frac{1}{2}\right)
    = \sum_\alpha \frac{1}{2}\hp_\alpha^2+\frac{\omega_\alpha^2}{2}\hq_\alpha^2,
\end{align}
where $\omega_\alpha$ is the photon frequency of mode $\alpha$ and $\hq_\alpha=\sqrt{\frac{1}{2\omega_\alpha}}(\ha_\alpha^\dag+\ha_\alpha)$. $a_\alpha^\dagger$ ($a_\alpha$) is the creation (annihilation) operator for photonic mode $\alpha$. The interaction between the cavity photon mode and the molecular system is described by
\begin{equation}\label{eq:hmc}
    \hH_{mc}=\sum_\alpha \omega_\alpha \hq_\alpha\blam_\alpha\cdot\hbu + \frac{(\blam_\alpha\cdot\hbu)^2}{2}.
\end{equation} 
$\blam_\alpha$ is the polarization vector of cavity mode $\alpha$. $\hbu$ is the total dipole operator, $\hbu= \sum_{IJ} \bu_{IJ} \hc^\dag_I \hc_J$, where includes both permanent dipole $\bu_{II}$ of each adiabatic state and the transition dipole $\bu_{IJ}$ between the adiabatic states $I$ and $J$. Dropping the last term of Eq.~\ref{eq:hmc} reduces it to the Rabi model, while omitting the last term and the counter-rotating terms leads to the Jaynes-Cummings (JC) model.~\cite{Taylor2020PRL}

In this work, we only consider the case with one photon mode so that the index $\alpha$ will be dropped, and we will denote the cavity frequency as $\omega_c$ and the cavity polarization vector as $\blam$. The molecule-cavity coupling term is defined as
\begin{align}
  \label{eqn:gij_vec}  
  g_{IJ}=\sqrt{\frac{\omega_c}{2}}\blam\cdot\bu_{IJ}.
\end{align}
If we assume the cavity polarization vector $\blam$ always to be aligned with the molecular transition dipole moment $\bu_{IJ}$, Eq.~\ref{eqn:gij_vec} can be reduced to $g_{IJ}=g|\bu_{IJ}|$,
where $g = \sqrt{\frac{\omega_c}{2}}|\blam|$ is a scalar parameter that is only related to the property of the cavity and independent of molecules.
We refer to this $g$ as the ``coupling strength'' in this work.
We only consider the transition dipole from the ground state ($I \equiv 0$) to the $J$th excited state, so index $I$ is omitted later.

With the adiabatic states calculated from any electronic structure methods, the polaritonic state can be readily computed by casting Eq.~\ref{eqn:eqham}into the matrix form within the product of electronic adiabatic state and photonic Fock states
\begin{align}
  \label{eqn:eqham_matrix}
    H_{Ii,Jj} = \braket{\phi_I, i | \hH | \phi_J, j}.
\end{align}
Here $\ket{\phi_I}$ is the wavefunction for the molecular excited state $I$, and $\ket{i}$ is the wavefunction for photonic excitation $i$. As a result, the polariton states can be expressed as 
\begin{align}
  \label{eqn:plariton_wf} 
  \ket{\Phi_N} = \sum_{Ii}A_{Ii}^N\ket{\phi_I,i},
\end{align}
where $N$ denotes the $N$th polariton states and $A_{I\alpha}^K$ is the coefficient that can be obtained by diagonalizing Eq.~\ref{eqn:eqham_matrix}. The polariton states are formed by the superposition of the molecular states with the photonic state.

\subsection{Ab initio multiple cloning for nonadiabatic polariton dynamics}\label{sec:aimc}

When multiple molecules are coupled to the cavity mode, the polariton states are delocalized among them, competing with the thermal fluctuation-induced localization. It is thus crucial to accurately describe the decoherence process to simulate exciton-polariton dynamics correctly. Common MQC methods lack decoherence. For example, the ad-hoc decoherence correction is usually introduced in the TSH algorithms~\cite{Nelson2020CR}. Ehrenfest dynamics, on the other hand, overestimate coherence due to its mean-field nature. The semiclassical AIMC method, based on multiconfigurational Ehrenfest dynamics augmented by wavepacket cloning, has been shown to naturally incorporate decoherence effects~\cite{Song2021JCTC, Freixas2021JPCL}. Consequently, this technique is more suitable for simulating the dynamics inside an optical cavity. We will briefly review the AIMC method and introduce its extension to polariton problems. 

\subsubsection{Multiconfigurational Ehrenfest Dynamics}\label{eq_mce}
Multiconfigurational Ehrenfest (MCE)~\cite{Makhov2017CP} is a generalization of the well-known Ehrenfest formalism.~\cite{Billing1983CPL} Under the MCE framework, the wavefunction is a linear combination of Ehrenfest configurations, and each configuration moves along its own Ehrenfest (mean-field) trajectory. Mathematically, the molecular wavefunction $\ket{\Psi}$ is expressed in the trajectory-guided Gaussian basis function (TBF) representation ($\ket{\psi_n}$),
\begin{align}
  \label{eqn:mcewf}
  \ket{\Psi(t)}=\sum_n c_n \ket{\psi_n(t)},
\end{align}
where $\ket{\psi_n(t)}$ is the wavefunction for trajectory $n$ and $c_n$ is the corresponding coefficient.
$\ket{\psi_n(t)}$ can again be expressed as a linear combination of adiabatic basis $\ket{\phi^{(n)}_I}$ as
\begin{align}
  \label{eqn:ehrwf}
  \ket{\psi_n(t)} = \ket{\chi_n(t)} \sum_I a^{(n)}_I \ket{\phi^{(n)}_I(t)},
\end{align}
where $\ket{\chi_n(t)}$ is a Gaussian coherent state moving along the trajectory.
The couplings between TBFs in the MCE approach are described by the equations of motion of $c_n(t)$, which can be readily obtained by substituting Eq.~\ref{eqn:mcewf} into the Schr\"odinger equation, yielding
\begin{align}
    i\hbar\sum_n S_{mn} \dot{c}_n = 
    \sum_n \left[H_{mn} - i\hbar\braket{\psi_m | \frac{d\psi_n}{dt}} \right]c_n,
\end{align}
where $m$ and $n$ are the indices for the TBFs.
The overlap matrix $S_{mn}$, including both the nuclear part and electronic part, is
\begin{align}
  \label{eqn:s}
    S_{mn}=\braket{\psi_m | \psi_n}
    =\braket{\chi_m | \chi_n}\sum_{I,J}a^{(m)*}_Ia^{(n)}_J
    \braket{\phi^{(m)}_I|\phi^{(n)}_J},
\end{align}
where $\ket{\phi^n_J}$ is molecular excited state $J$ for trajectory $n$.
The $H_{mn}$ is the matrix element of the full Hamiltonian, written as 
\begin{align}
  \label{eq_hmn}
  H_{mn}=\sum_{I,J}a^{(m)*}_Ia^{(n)}_J \bra{\chi_m\phi^{(m)}_I}T+V\ket{\chi_n\phi^{(n)}_J},
\end{align}

For polariton, one can replace the molecular wavefunction with the polariton state wavefunction in the expressions above. For example, by inserting Eq.~\ref{eqn:plariton_wf} into Eq.~\ref{eqn:s}, we can obtain the overlap in the presence of the cavity mode as,
\begin{align}
    S_{mn} =& \braket{\chi_m|\chi_n}\sum_{N,M}a^{(m)*}_Na^{(n)}_M
    \braket{\Phi^{(m)}_N | \Phi^{(n)}_M} \nonumber\\
    =& \braket{\chi_m|\chi_n} \sum_{N,M} a^{(m)*}_Na^{(n)}_M
    \sum_{I,J,i} (A_{Ii}^N)^*A_{Ji}^M\braket{\phi^{(m)}_I|\phi^{(n)}_J},
\end{align}
where $N$ and $M$ are the indices for polariton states.
The other quantities can be re-expressed similarly as well for polariton states.~\cite{Zhang2019JCP}

The nuclear kinetic energy part in Eq.~\ref{eq_hmn} can be expressed as
\begin{align}
  \label{eqn:kinetic}
  \bra{\chi_m\phi^{(m)}_I}T\ket{\chi_n\phi^{(n)}_J}= &
  -\braket{\chi_m | \frac{\hbar^2}{2}\nabla_{\bR}\bM^{-1}\nabla_{\bR} | \chi_n} 
  \nonumber\\
  & \times
  \langle\phi^{(m)}_I\ket{\phi^{(n)}_J},
\end{align}
where $\bM^{-1}$ is the inverse of the mass matrix and $\nabla_\bR$ are the gradients with respect to the atomic coordinates $\bR$.
Eq.~\ref{eqn:kinetic} can easily be integrated analytically.~\cite{Makhov2014JCP}
Follow the previous derivations,~\cite{Makhov2014JCP, Freixas2018PCCP, Song2021JCTC} The potential energy matrix elements, on the other hand, are approximated using a bra-ket averaged Taylor expansion,
\begin{align}
  &\bra{\chi_m\phi^{(m)}_I}V\ket{\chi_n\phi^{(n)}_J}=\frac{1}{2}
    \braket{\phi^{(m)}_I | \phi^{(n)}_J} \braket{\chi_m | \chi_n} \times \nonumber\\
  &\left\{(V^{(m)}_I+V^{(n)}_J) + \frac{i}{4\alpha\hbar}(\bP_n-\bP_m)\cdot(\nabla_{\bR}V^{(m)}_I+\nabla_{\bR}V^{(n)}_J)\right . \nonumber\\
  &\left . -\frac{1}{2}(\bR_m-\bR_n)\cdot(\nabla_{\bR}V^{(m)}_I-\nabla_{\bR}V^{(n)}_J)\right\},
\end{align}
where $\bR_n$ and $\bP_n$ are the phase space centers of the $n$th TBF, $\alpha$ is a width parameter of $\ket{\chi_n(t)}$ whose value can be chosen based on Ref.~\citenum{Thompson2010CP}, and $V^{(n)}_I$ is the potential energy surface for the $I$th excited state of $n$th TBF.

\subsubsection{ AIMC Algorithm}\label{subsec:aimc}
The ab initio multiple cloning (AIMC) method advances MCE methods by introducing a cloning procedure for the wavepackets to avoid exploration of unphysical space with TBFs. Similar to the MCE method, the individual TBFs of AIMC follow Ehrenfest EOM. However, the basis set is expanded when these TBFs become sufficiently mixed.~\cite{Makhov2014JCP, Makhov2015PCCP, Makhov2016FD} Consequently, AIMC avoids prolonged evolution on the mean-field PES. Previous works~\cite{BenNun1998, BenNun2007} have shown that with this basis set expansion technique, AIMS and AIMC can converge to the exact solution of the Schr\"odinger equation for a large enough basis set. 

Following the wavefunction expression (Eqs.~\ref{eqn:mcewf}-\ref{eqn:ehrwf}) in the MCE formalism, when multiple PESs become different in shape, which usually occurs near electronic level crossings, the original basis of TBFs in AIMC is expanded via ``cloning'' one TBF into two copies,
\begin{align}
  \ket{\psi_{n_1}} =& \ket{\chi_n}\left(\frac{a^{(n)}_I}{|a^{(n)}_I|} \times \ket{\phi^{(n)}_I} + \sum_{J\ne I} 0 \times \ket{\phi^{(n)}_J} \right)   \nonumber\\
  \ket{\psi_{n_1}} =& \ket{\chi_n}\left(0 \times \ket{\phi^{(n)}_I} + \frac{1}{\sqrt{1 - |a^{(n)}_I|^2}}\sum_{J\ne I} a^{(n)}_J\ket{\phi^{(n)}_J} \right).  
\end{align}
During this cloning process, one of the clones $\ket{\psi_{n_1}}$ is in a pure state, and the other clone $\ket{\psi_{n_2}}$ combines contributions from all other electronic states, with all coefficients being renormalized. The corresponding MCE amplitudes are given by
\begin{align}
  c_{n_1} =& c_n a^{(n)}_I \nonumber\\
  c_{n_2} =& c_n \sqrt{1 - |a^{(n)}_I|^2}.
\end{align}
As a result, the total contributions of the two clones $\ket{\psi_{n_1}}$ and $\ket{\psi_{n_2}}$ to the total wavefunction in Eq.~\ref{eqn:mcewf} remains conversed,
\begin{align}
  c_n\ket{\psi_n} = c_{n_1}\ket{\psi_{n_1}} + c_{n_2}\ket{\psi_{n_2}}.
\end{align}
Through cloning, the Ehrenfest method's mixed states are represented by superpositions of individual TBFs to avoid prolonged mixed state propagation that is known to ``over-cohere'' and violate the detailed balance.~\cite{Makhov2014JCP} One major problem of the cloning approach is that the computational cost scales at least linearly with respect to the number of clones.~\cite{Curchod2018} As a result, the number of clones in a simulation must be limited to avoid an intractable increase in numerical cost. The NEXMD package currently implements 3 criteria are implemented to determine if a trajectory will be cloned. These criteria have been discussed extensively in previous works~\cite {Freixas2018PCCP}.
With these 3 criteria, one can limit the total number of clones to a reasonable number, avoiding overwhelming computational costs, and provide sufficient expansion of the basis set to mimic the bifurcation of the wavefunctions due to nonadiabatic processes. Note that multiple clones can be created in one step.
After a cloning event, if all criteria are still met for the TBF with a mixed state, a new clone will be created from this basis function.
More details of the AIMC algorithm implementation in the NEXMD package can be found in Refs.~\cite {Song2020JCTC} and \citenum{Freixas2018PCCP}. 

\section{Polaritonic state relaxation of pyridine}\label{sec:numerical}
\subsection{Simulation details}\label{sec:details}
The previously studied pyridine molecule is used to demonstrate the MCE-AIMC method for polariton chemistry to compare the relaxation dynamics with and without light-matter interactions. The maximum number of clones is limited to 16. In this work, one pyridine molecule is coupled to a single photon mode. The  Austin model 1 (AM1) SQM model~\cite{Dewar1985} is employed to calculate the electronic structure of pyridine.
After initial equilibrium, we ran a 100,000-step ground-state Born-Oppenheimer molecular dynamics simulation to sample conformations for the excited state calculations and initial conditions for subsequent NAMD simulations.
The time step is set to 0.1 fs, and each geometry snapshot is taken every 100 steps (10 fs), resulting in 1,000 initial configurations in total for polariton simulations.
We then perform NAMD simulations at 100 K and 300 K by employing both AIMC and TSH algorithms. The cavity photon frequency $\omega_c$ is set to be 4.33 eV, matching the second absorption peak in the UV spectrum (see Fig.~\ref{fig:spectrum}). The light-matter coupling strength parameter $g$ is chosen to be 0, 0.1, and 0.2~eV to mimic different regimes. The initial excitation state of the pyridine molecule in each NAMD trajectory is set to S$_3$ for no cavity case and S$_4$ with the cavity, which corresponds to the same adiabatic state of the molecule. 112 and 500 trajectories are employed in each AIMC and TSH simulation, respectively.

\subsection{Results and discussions}
\begin{figure}[!htb]
  \centering
  \begin{minipage}[t]{1.0\linewidth}
    \centering
    \includegraphics[width=1.0\linewidth]{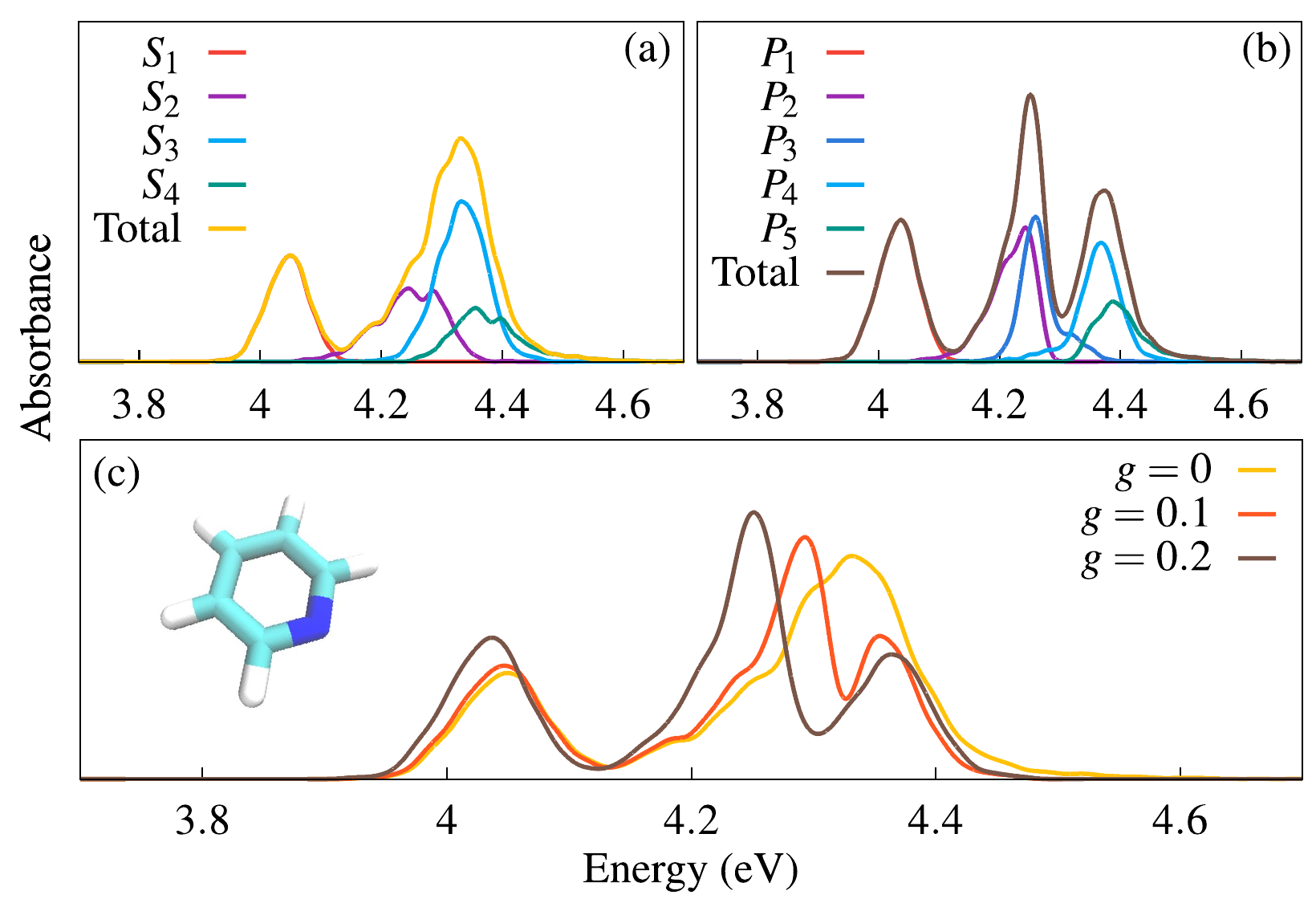}
    \caption{Optical absorption spectrum of pyridine.
    (a) The contribution of the first four excited states (S$_1$ - S$_4$) of pyridine and the net spectrum.
    (b) Polariton spectrum at 100 K with the coupling strength $g = 0.2$.
    $P$ denotes polariton states.
    The photon frequency $\omega_c = 4.33$ eV resonates with the second peak in panel (a). 
    Note that in the presence of the cavity mode, the single peak at 4.33 eV splits into two peaks, corresponding to the lower (with lower energies) and upper (with higher energies) polaritons, respectively. The energy gap between the two states is the Rabi splitting.
    (c) Optical spectra of pyridine at 3 different light-matter coupling strengths $g$ showing the polariton states.
    The Rabi Splitting increases as the light-matter coupling strength increases. However, in this specific case, increased coupling strength has a small impact on the energy of the upper polariton, but the lower polariton, on the other hand, is pushed to the lower energy region.}
    \label{fig:spectrum}
  \end{minipage}
\end{figure}

Figure~\ref{fig:spectrum} shows the NEXMD optical absorption spectra that include the first four excited states of pyridine with (a) and without (b) the cavity mode simulated at ambient conditions. Without the cavity mode, there are two prominent peaks in the spectrum. S$_1$ is energetically lower, forming the first peak, while S$_2$-S$_4$ states are relatively closer to each other, forming the second peak at 4.33 eV. Note that among these 3 states, S$_3$ has the strongest absorption and centers around 4.33 eV. As a result, when the molecule is photoexcited at 4.33 eV, S$_3$ will be the dominant initial state among the trajectories.

Figure~\ref{fig:spectrum}b shows the polariton spectrum of all 5 polariton states when $g = 0.2$ at a photon frequency of 4.33 eV. When the pyridine molecule is coupled to the cavity mode, the single peak in the bare molecule case (orange line) is split into two peaks, with the one with lower (higher) energy representing the lower (upper) polariton. The energy gap between the two peaks is the so-called Rabi splitting, which indicates the strong coupling between the molecule and the cavity mode. Due to the disorder of the geometries induced by thermal fluctuations, the splitting is not symmetric, with the lower polaritons exhibiting stronger absorption.

Figure~\ref{fig:spectrum}c shows the Rabi splitting with respect to different coupling strengths, with a photon frequency of 4.33 eV. As the coupling strength $g$ increases, the Rabi splitting also increases. In this case, the larger splitting mainly results from the lower polariton's redshift, with the upper polariton's energy almost unchanged. As polariton states are linear combinations of the molecular excited states and the photon excitations, when the photon contribution is low, the resulting polariton state will be close to the corresponding excited state. Even though the peak at 4.05 eV is energetically far from the photon frequency of 4.33 eV, the molecules are still coupled to the cavity mode, albeit with weaker coupling strength.

\begin{figure}[htb]
  \centering
  \begin{minipage}[t]{1.0\linewidth}
    \centering
    \includegraphics[width=1.0\linewidth]{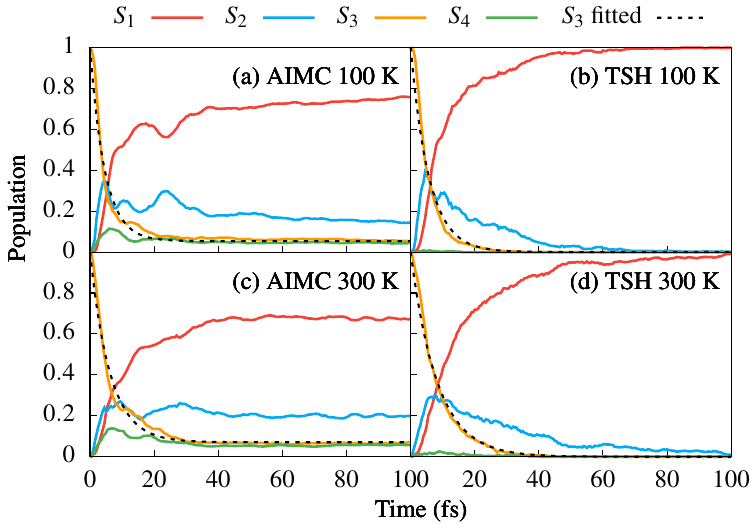}
    \caption{The populations of the first four excited states of pyridine computed from AIMC (left) and TSH (right) at 100 K and 300 K without cavity mode ($g=0$). The population for S$_3$ is fitted with an exponential function (black dashed lines). The oscillations in the populations (for example, S$_2$) in the AIMC simulations originate from the decoherence process, highlighting the importance of a rigorous treatment of decoherence. Note that the so-called ``classical populations'' are plotted here for TSH. The population for $k$th state at time $t$ is defined as $p_k(t)=N_k(t)/N_{traj}$, where $N_k(t)$ is the number of trajectories in the $k$th state and $N_{traj}$ is the total number of trajectories.}
    \label{fig:aimc}
  \end{minipage}
\end{figure}

Figure~\ref{fig:aimc} shows the time evolution of the populations of the first four excited states (no cavity involved) with respect to time for AIMC simulations (left column) and TSH (right column) simulations at 100 K and 300 K. The black dashed lines are exponential fittings of the population of S$_3$. The population decay of the initially excited state and build-up of the final state ($S_1$) can be fitted with two exponential functions, $P_i(t) = a \cdot \frac{[1+\cos^2(\omega t)] }{2}e^{-t/T} + P_i(\infty)$ and $P_f(t)= P_f(\infty)\{1 - \frac{[1+\cos^2(\omega t)]}{2}e^{-t/\tau}\}$, respectively. Here, $P_{i/f}(\infty)$ are the final populations of the initially excited state and final state at the long time limit, respectively. $\cos^2(\omega t)$ function is introduced to account for the initial coherent oscillations. And $c$ is the final population of $S_1$. Then, the population depletion time $T$ and relaxation time $\tau$ are fitted against the simulation result. The fitted values of $T$ and $\tau$ for all the simulations are listed in Table.~\ref{tbl:rate} and Table.~\ref{tbl:relaxrate}, respectively. During the initial dynamics stage, oscillatory population behavior for all states included in the AIMC simulations can be observed. Such behavior typically originates from coherent vibronic dynamics.~\cite{Song2021JCTC} These oscillations are less evident in the TSH simulations, indicating an insufficient treatment of coherence.
Besides, the difference between AIMC and TSH is also evident in the residual population of higher states at the long-time limit.
At 100 fs, S$_3$ and S$_4$ in AIMC simulations still have substantial populations, but their populations already approach 0 for TSH simulations. At higher temperatures, the depletion time for both AIMC and TSH becomes slower. This counter-intuitive behavior can be understood by examining the initial excitation energy distribution shown in Fig.~\ref{fig:energy} in the Appendix. At 100 K, the initial configurations are more localized, so the average initial excitation energy is slightly higher than 300 K for both simulations, compensating for the initial kinetic energy difference.

\begin{table}[!htb]
  \centering
  \caption{Fitted population depletion time constants for the population decrease of $S_3$ (no cavity case) or $S_4$ (with cavity) for all simulations.
  }
  \begin{tabular*}{0.45\textwidth}{@{\extracolsep{\fill}}cc|ccc}
    \hline
    \hline
    \footnotesize
    &      & \multicolumn{3}{c}{Population depletion time (fs)} \\
    &      & $g=0$      & $g=0.1$~eV      & $g=0.2$~eV      \\ \hline
    \multicolumn{1}{c|}{\multirow{2}{*}{100 K\ \ }} & AIMC\ \ \ \ \ \ & 4.59       & 4.60         & 2.63         \\
    \multicolumn{1}{c|}{}                       & TSH\ \ \ \ \ \    & 5.70       & 4.81         & 4.61         \\
    \hline
    \multicolumn{1}{c|}{\multirow{2}{*}{300 K\ \ }} & AIMC\ \ \ \ \ \  & 6.18       & 5.19         & 3.98         \\
    \multicolumn{1}{c|}{}                       & TSH\ \ \ \ \ \   & 8.25       & 6.30         & 5.54         \\ \hline
  \end{tabular*}%
  \label{tbl:rate}
\end{table}

\begin{figure}[htb]
    \centering
    \includegraphics[width=0.5\textwidth]{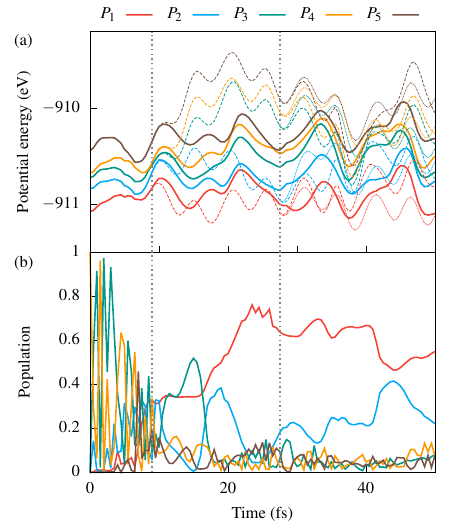}
    \caption{A typical AIMC trajectory is visualized with all 5 polariton states, denoted by $P$. Note that other trajectories can behave very differently. The initial trajectory (solid) is cloned at 9 fs (dashed lines). At 41 fs (dotted lines), the second trajectory creates its own clone. The two grey lines mark the time when the clones are created. (a) Time evolution of the potential energy of all 5 states for each TBF. (b) Time evolution of the total excited state population. Note that the population is a weighted average of all TBFs. The population of each trajectory is not necessarily the same as the overall population plotted here.}
    \label{fig:clone}
\end{figure}

Figure~\ref{fig:clone}a and b show the potential profiles and excited-state populations of a typical AIMC trajectory with cloning events clearly visible. This AIMC trajectory is simulated at 300 K with the light-matter coupling strength $g = 0.2$~eV. The second TBF (first clone) is created at 9 fs. At this moment, the most populated states $P_3$ and $P_1$ are diverging from each other, where $P$ refers to the polariton states. As a result, the mean-field force is no longer a good representation of the whole system, and a new TBF has been created to address this issue. ~\cite{Freixas2018PCCP, Song2021JCTC} Note that, from this point on, there is more than one trajectory in the simulation, so the overall population, shown in Fig.~\ref{fig:clone}b, is a weighted average among the contribution of all trajectories. The most populated states of the overall population are not necessarily the same as those of the individual trajectories. The cloning criteria, on the other hand, are always about the most populated states within each trajectory.
Consequently, it is possible that states involved in a cloning event do not match the overall populations shown in Fig.~\ref{fig:clone}b. It is necessary to examine per-trajectory populations or adiabatic state coefficients. Then, at 27.5 fs, the most populated state of the cloned TBF $P_1$ diverges from the mean field, causing the second clone (third TBF) to be created. The PES plots at 27.5 fs clearly show that the dotted lines, representing the second clone, originate from the first clone (dashed lines) rather than the original trajectory. This highlights how the cloning mechanism works in AIMC. Every trajectory can create its own clones, and when necessary, multiple clones can be created at one time step, as long as the cloning criteria are met and the total number of clones is within the user-defined limit. As we have explained, this limit exists due to the consideration from the computational cost perspective.

\begin{table}[!htb]
  \centering
  \caption{Fitted relaxation time for the population build-up of $S_1$ due to the relaxation of upper states.
  }
  \begin{tabular*}{0.45\textwidth}{@{\extracolsep{\fill}}cc|ccc}
    \hline
    \hline
    \footnotesize
    &      & \multicolumn{3}{c}{Relaxation time (fs)} \\
    &      & $g=0$      & $g=0.1$~eV      & $g=0.2 $~eV      \\ \hline
    \multicolumn{1}{c|}{\multirow{2}{*}{100 K\ \ }} & 
    AIMC\ \ \ \ \ \ &  13.91  &  28.24  &  20.27  \\
    \multicolumn{1}{c|}{}  &
    TSH\ \ \ \ \ \  & 16.74   & 25.89   & 23.02   \\
    \hline
    \multicolumn{1}{c|}{\multirow{2}{*}{300 K\ \ }} 
    & AIMC\ \ \ \ \ \  & 14.12 & 31.71 & 25.38 \\
    \multicolumn{1}{c|}{}
    & TSH\ \ \ \ \ \   & 16.74 & 25.88  & 23.03 \\ \hline
  \end{tabular*}%
  \label{tbl:relaxrate}
\end{table}

\begin{figure}[htb]
  \centering
  \begin{minipage}[t]{1.0\linewidth}
    \centering
    \includegraphics[width=0.99\linewidth]{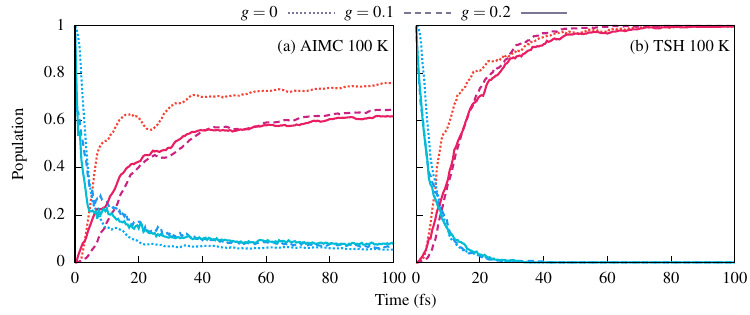}
    \caption{Population evolution of S$_1$ (red) and the initial excited states, S$_3$ without cavity and S$_4$ with cavity, (blue) with respect to 3 different light-matter coupling strength $g$ for AIMC (left) and TSH (right) simulated at 100 K. As the coupling strength increases, the equilibrium population of S$_1$ decreases for AIMC simulations.
    For TSH, on the other hand, the coupling strength has negligible influence on the populations beyond transient time.}
    \label{fig:coupling}
  \end{minipage}
\end{figure}

Figure~\ref{fig:coupling} shows the populations as a function of time simulated with different light-matter coupling strengths at 100 K. Two states, P$_1$ in red and the initially excited state (S$_3$ without cavity and P$_4$ with cavity) in blue, are shown. As Fig.~\ref{fig:spectrum}b shows, the cavity mode pushes up the energy level of the upper polariton. Therefore, the energy difference between P$_4$ and P$_1$ is higher than that between S$_3$ and S$_1$ in the absence of the cavity. From an energetic perspective, one would expect the depletion time of polaritonic states to increase. However, Fig.~\ref{fig:coupling}a and Table~\ref{tbl:rate} show that the coupling to the cavity reduces the depletion time in the AIMC simulations. Because all excited states are coupled to the same cavity mode, additional population transfer pathways between states are enabled. The depletion of the initial excited state will be facilitated when the enhanced transfer suppresses the negative impact of the increased energy gap. This explanation is also evidenced by comparing $g = 0.1$~eV and $g = 0.2$~eV. When the coupling strength is doubled, though the energy of P$_4$ remains almost unchanged, the interactions between the excited states are greatly enhanced. As a result, the depletion time in the AIMC simulation is significantly faster, 2.63 fs for $g = 0.2$~eV and 4.6 fs for $g = 0.1$~eV (see Table~\ref{tbl:rate}). In contrast, the relaxation time $\tau$, the time that the wavepacket decays to $S_1$ (or $P_1$) state as shown in Figure~\ref{fig:coupling} and Table~\ref{tbl:relaxrate}, is increased. This is because the coupling to the cavity introduces coherent energy transfer among the polaritonic states, leading to slower relaxation. 
Even though the depletion time for the TSH simulation is decreased as well, the difference is rather subtle, shown in Fig.~\ref{fig:coupling}b. We believe a similar cavity-mode-mediated population transfer mechanism also exists in the TSH simulations, but the instantaneous decoherence approach essentially cancels out its impact on the dynamics. 

Compared to the TSH results, AIMC results feature more oscillations in the initial relaxation dynamics. As previously discussed, this oscillatory behavior originates from the vibronic coherent dynamics. The nonadiabatic force accounts for the vibration-induced electronic population exchange.~\cite{Freixas2021JPCL} However, this coherent behavior is not observed in the TSH simulations as was reported in previous NEXND simulations of other molecular systems~\cite{Nelson2020CR}. The light-matter coupled system is expected to enhance this vibronic population transfer when the molecule is strongly coupled to a cavity mode due to the interactions between excited states and the photon excitation. As a result, the oscillations in the initial stage are further enhanced in polariton dynamics (the dashed and solid lines in Fig.~\ref{fig:coupling}a). On the other hand, TSH does not capture such oscillations, with or without the cavity mode.

Another interesting observation is the final population. For TSH, though the depletion time is decreased with increasing light-matter coupling, the final populations of both states are not fundamentally affected. However, for AIMC, the equilibrium population of P$_1$ is significantly lower when there is light-matter coupling. AIMC simulations allow for transitions from lower to higher states, as all states included in the simulations are coupled to the same cavity mode. We conjecture that the cavity mode leads to more population transfer to higher states (for example, P$_2$) in AIMC simulations. Because the coupling to the cavity causes fast depletion of the population of the initially excited state and slower nonadiabatic relaxation due to the active exchange of populations in the presence of the cavity. In TSH, hops are forbidden, preventing population transfer to high states. The introduction of the ad-hoc decoherence correction overcomes this issue but completely removes the populations in high-energy states. 
{Since this behavior (population transfer to higher states) is completely missing in TSH, there is no enhancement at all}, such that the existence of the cavity mode does not affect the populations of different states.

\section{Conclusion}\label{summary}
This study presents a computational framework for modeling molecular polariton dynamics based on the semiclassical ab initio multiple cloning (AIMC) algorithm implemented in the NEXMD package. Calculations of energies, dipoles, gradients, and nonadiabatic couplings are performed in the strong light-matter interaction regime, allowing for efficient simulation of exciton polariton dynamics on the fly. To demonstrate and validate our implementation, we have modeled the photoinduced dynamics of the pyridine molecule subject to different temperatures and light-matter coupling strengths. The results are compared to the data obtained from the standard TSH modeling.
The effects of the cloning procedure are examined through the evolution of the excited state populations and induced vibrational motions coupled to the electronic dynamics. The AIMC approach leads to higher residual polaritons for higher excited states and longer relaxation times than that obtained with the TSH method.
The cavity mode further enhances the population transfer to the higher states and slows down the relaxation dynamics.
This highlights that the cavity mode can potentially enable new reaction pathways in photoinduced nonadiabatic dynamics, which can not be captured with the traditional surface hopping method. 

However, calculating the forces and nonadiabatic couplings for polariton states is computationally demanding in the presence of the cavity mode compared to the pristine case.
The derivative of the transition dipole moment is a necessary component for NAMD simulations of exciton-polariton. Due to numerical instabilities, calculating such a derivative is computationally expensive and suffers from convergence issues.~\cite{Zhang2019JCP}
Combined with the extra numerical cost of AIMC, simulating more complicated molecules or the scenario where many molecules are coupled to the cavity mode can be demanding with the present framework, even when deploying with relatively cheap semiempirical methods. One possible approach is to utilize a machine learning model.
Recently, we have presented a machine-learning framework~\cite{Li2024} where one can train the neural network with a dataset collected for one molecule. Subsequently, the energies, forces, dipoles, and nonadiabatic couplings for the same molecular species in different configurations can be simultaneously predicted using a neural network. This machine learning framework enables the simulation of exciton-polaritons in systems with hundreds of molecules. In our future studies, we plan to combine these data science techniques with the AIMC methodology to achieve the simulation of exciton-polaritons in the collective coupling regime.

\begin{acknowledgments}
We acknowledge the support from the US DOE, Office of Science, Basic Energy Sciences, Chemical Sciences, Geosciences, and Biosciences Division under Triad National Security, LLC (``Triad") contract Grant 89233218CNA000001 (FWP: LANLECF7). This research used computational resources provided by the Institutional Computing (IC) Program and the Darwin testbed at LANL, funded by the Computational Systems and Software Environments subprogram of LANL's Advanced Simulation and Computing program. This work was partly performed at the Center for Integrated Nanotechnologies (CINT), a US Department of Energy, Office of Science user facility at LANL. 
LANL is operated by Triad National Security, LLC, for the National Nuclear Security Administration of the US Department of Energy (Contract No. 89233218CNA000001).

\end{acknowledgments}

\bibliography{ref}

\appendix

\setcounter{figure}{0}
\renewcommand\thefigure{S\arabic{figure}}

\section{Nonadiabatic coupling vector for polariton states}

For a pair of pure electronic excited states $I$ and $J$, the nonadiabatic couplings (NAC) between them can be written as
\begin{align}
  \bd_{IJ} = \braket{\phi_I | \frac{\partial \phi_J}{\partial \bR}}.
\end{align}
Now replacing the electronically excited state wavefunction $\phi_I$ with the polariton state wavefunction $\Phi_N$ (Eq.~\ref{eqn:plariton_wf}), one can obtain the NAC between two polariton states $N$ and $M$,
\begin{align}
  \bd_{NM} =& \braket{\Phi_N | \frac{\partial \Phi_M}{\partial \bR}} \nonumber \\
  =& \sum_{I,i} (A_{Ii}^N)^*\bra{\phi_I,i} \frac{\partial}{\partial \bR} \sum_{J,j}A_{Jj}^M\ket{\phi_J,j} \nonumber \\
  =& \sum_{I,J,i}(A_{Ii}^N)^*A_{Ji}^M \braket{\phi_I | \frac{\partial \phi_J}{\partial \bR}} \nonumber \\
  =& \sum_{I,J,i}(A_{Ii}^N)^*A_{Ji}^M\bd_{IJ}.
\end{align}
The second Fock state index $j$ is gone because $\braket{i | j} = \delta_{ij}$. Therefore, obtaining the corresponding quantity for polariton states is straightforward once the NAC between electronic adiabatic states is known.

\section{Initial energy distribution}
\begin{figure}[H]
  \centering
  \begin{minipage}[t]{1.0\linewidth}
    \centering
    \includegraphics[width=0.9\linewidth]{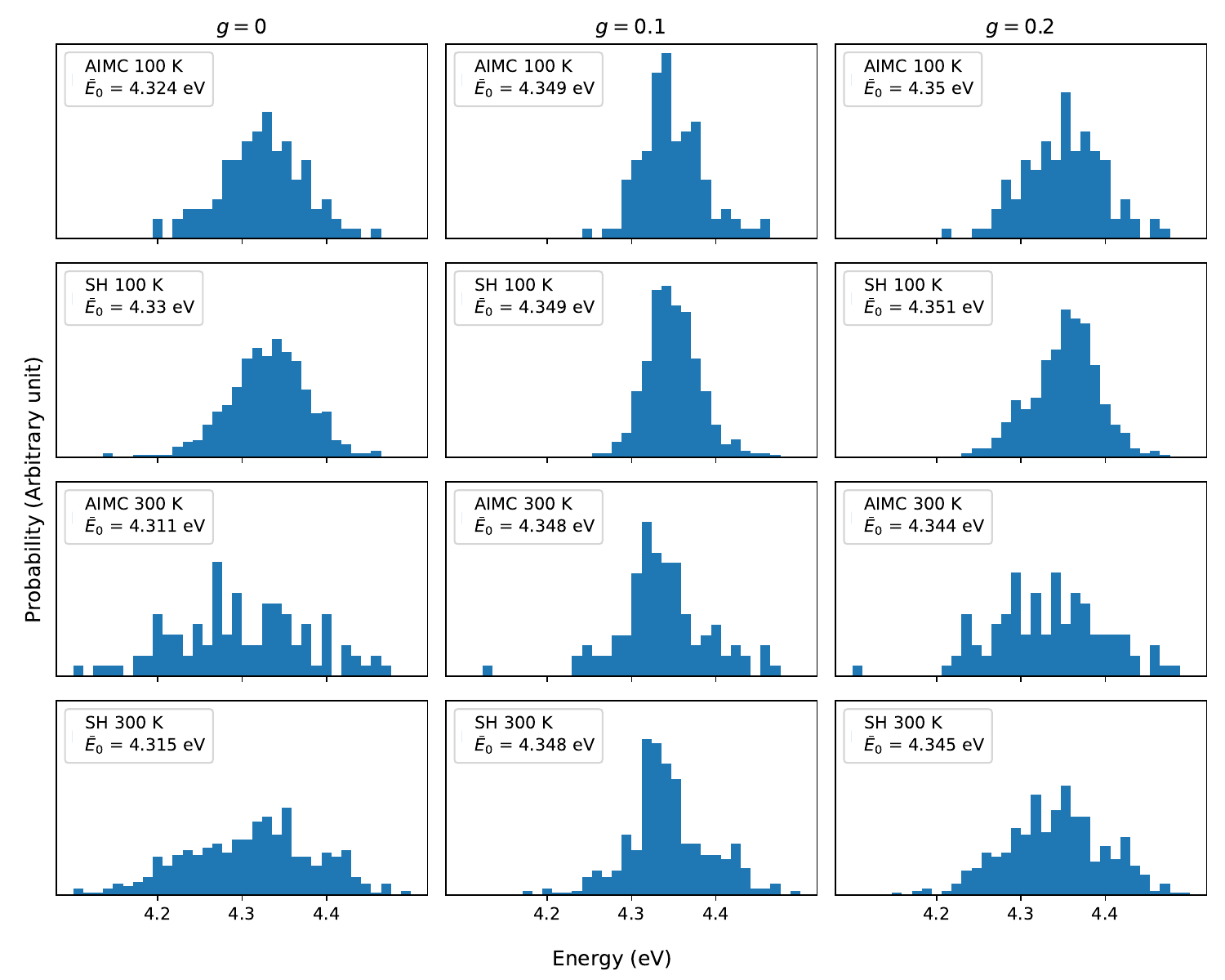}
    \caption{Initial excitation energy distributions of all simulations with average excitation energies shown in the legend.
    When there is no cavity ($g = 0$), all molecules are initially excited to S$_3$, corresponding to the peak at 4.33 eV in Fig.~1 in the main text.
    With cavity ($g = 0.1$ or $g = 0.2$), the molecular excited states hybridize with the photon mode.
    All molecules are excited to S$_4$, which forms the upper polariton in the spectrum.
    In general, simulations at 100 K have slightly higher average initial excitation energy due to more localized geometries.
    Different light-matter coupling strengths do not greatly impact the upper polariton state, though the energy of the lower polariton is pushed down (shown in Fig.~1). }
    \label{fig:energy}
  \end{minipage}
\end{figure}
\end{document}